# WEB ENGINEERING

Yogesh Deshpande[1], San Murugesan[2], Athula Ginige[1], Steve Hansen[1], Daniel Schwabe[3],
Martin Gaedke[4], Bebo White[5]

*(1) School of Computing and Information Technology, University of Western Sydney*
*Campbelltown Campus, Building 17, Locked Bag 1797*
*Penrith South DC NSW 1797, Australia*
*Email: {y.deshpande, a.ginige, s.Hansen}@uws.edu.au*

*(2) Southern Cross University, Coffs Harbour Campus, Hogbin Drive, Coffs Harbour, NSW 2457,*
*Australia*

*(3) Departamento de Informática. PUC-Rio, Rio de Janeiro, Brazil*
*Email: schwabe@inf.puc-rio.br*

*(4) Institute of Telematics, University of Karlsruhe, Postfach 6980, Zirkel 2, 76128 Karlsruhe*
*Germany*
*Email: gaedke@tm.uni-karlsruhe.de*

*(5) SLAC, 2575 Sand Hill Road, Menlo Park, CA 90425, USA*
*bebo@SLAC.Stanford.edu*

Web Engineering is the application of systematic, disciplined and quantifiable
approaches to development, operation, and maintenance of Web-based applications.
It is both a pro-active approach and a growing collection of theoretical and empirical
research in Web application development. This paper gives an overview of Web
Engineering by addressing the questions: a) why is it needed?  b) what is its domain
of operation? c) how does it help and what should it do to improve Web application
development? and d) how should it be incorporated in education and training? The
paper discusses the significant differences that exist between Web applications and
conventional software, the taxonomy of Web applications, the progress made so far
and the research issues and experience of creating a specialisation at the master's
level. The paper reaches a conclusion that Web Engineering at this stage is a moving
target since Web technologies are constantly evolving, making new types of
applications possible, which in turn may require innovations in how they are built,
deployed and maintained.

## 1   Introduction

Web Engineering is the application of systematic, disciplined and quantifiable approaches to development, operation, and maintenance of Web-based applications[1]. It is a response to the early, chaotic development of Web sites and applications as well as recognition of a divide between Web developers and conventional software developers[2, 3]. Viewed broadly, Web Engineering is both a conscious and pro-active approach and a growing collection of theoretical and empirical research. Special issues of journals[4, 5, 6, 7], an edited book of papers[8], series of workshops, tutorials and special tracks at international conferences (WWW7 - WWW 2003, HICSS 1999 - 2001, SEKE'02, SEKE'03 and others), and dedicated international conferences (ICWE2002, ICWE2003) attest to the level of activity in this field.  However, this list is only a partial representation of the work undertaken in the field and the experiences of the multitude of Web developers.  The practice, good and bad, is leading the theory, to quote a remark made about the field of software maintenance a few years ago[9].

This paper is the first in a series of papers on Web Engineering.  It gives an overview of Web Engineering. It is not a comprehensive review of the work published so far although it necessarily draws upon contributions from researchers and practitioners across the world.  Other papers in the series, to be published in the future issues of the Journal, will cover various topics in greater detail.

A note on terminology: the literature variously refers to Web sites, Web-based applications, Web-based systems, Web applications and other variants of these when discussing Web Engineering. This paper will use the term Web applications to represent all the variations. In any case, we will consider sites (or Web applications) that have some informational purpose, that help people perform some task. Further, for the sake of brevity, the term 'Web development' will be used as a short form to signify the development, deployment and maintenance of Web applications.

The paper is organised as follows.  Section 2 sets out the need for or the 'why' of Web Engineering, an issue tackled by several people, based on practice.  Section 3 briefly updates the taxonomy of Web applications, i.e. the domains in which Web Engineering operates.  In other words, 'what' constitutes Web Engineering.  It also elaborates upon the growing complexity of Web applications. Section 4 discusses the issues of developing, testing and maintaining Web applications, i.e. 'how' Web Engineering is doing what it should do, and indications for future work, in a given domain. Since Web Engineering is building a body of work, this section covers both practice and research questions. Section 5 deals with education and training of Web engineers. Web technologies are constantly evolving, making new types of applications possible, which in turn may require innovations in how they are built, deployed and maintained. Section 6 concludes the paper, suggesting that Web Engineering at this stage is still a moving target.

## 2   Need for Web Engineering

The need for Web Engineering is felt (or dismissed) according to perceptions of the developers and managers, their experiences in creating applications made feasible by the new technologies, and the complexity of Web applications.  In the early stages of Web development, White[10] and Powell[11], identified and emphasized  the need for engineering as in Web Document Engineering and Web Site Engineering.  Web Engineering, more generally, explicitly recognises the fact that good Web development requires multidisciplinary efforts and does not fit neatly into any of the existing disciplines.

## 2.1    Perceptions of Web Development

Web development is perceived at different levels, shown in Figure 1.

| |
|---|
| 6. Web project planning and management |
| 5. Web-based System |
| 4. Web Site Construction |
| 3. Web Site Design |
| 2. Web Page Design |
| 1. Web Page Construction |

Figure 1: Levels of perception in Web Development

For someone relatively new to Web development, be they developers, users or managers, the Web is manifested through the Web pages, the outcome of the simplest and most visible level (level 1 of Figure 1, above). It also happens to be the easiest to understand and master since it is built upon a mark-up language (HTML) rather than a programming language. The next level, Web Page Design, becomes apparent as the developers and managers gain experience.  If they are from Information Technology (IT) background they realise that special skills are required, many outside computer science itself, the background of software engineers. The non-IT managers and developers, on the other hand, may not start to appreciate the crucial role of programming, databases, networks and other IT areas till later. The page design, though, may not be regarded as problematic since there are many packages that promise to ease the burden of page design.  In software engineering terms, these two levels correspond to user interface, generally regarded as a matter of detail and lying more in human-computer interaction (HCI) arena.  The next level of perception regards Web Site Design, or Information Architecture for some. Here, the hypertextual nature of the Web comes into play, since good web sites provide good navigation structures (i.e, structures that help its users achieve their goals). This level has not been addressed at all by traditional software engineering, and again may involve skills outside computer science. In figure 1, only levels 4 to 6 deal with processes of interest to software engineers[12].

To add to the perceptual difficulties here, a large number of organisations enter the Web development at stage 3, i.e. by decreeing that they must have a 'Web presence'. Consequently, Web development may be viewed mainly in terms of "publishing" or "brand building/reinforcement", where lessons learnt from software engineering are regarded as irrelevant or simply ignored. The understanding and importance of other stages become clearer only after a Web site is created, and the realization that it is, after all, an information system.  The need for systematic, measurable and repeatable development processes then becomes apparent.  Late recognition of the importance of Web Engineering could then lead to a redesign and re-engineering of the existing sites and applications, resulting in wasted efforts and resources.

Thus, software engineering is applicable and necessary at the application and project management levels but is not sufficient for all the activities as depicted in figure 1.  Further, there is a consensus, explained below, that even where software engineering is applicable, more and newer development, testing and maintenance methods will have to be found to deal with specific problems of Web development.

## 2.2    Web Developers' Experience, New Technologies and Expert Consensus

The need for Web Engineering has been debated and discussed in several fora, including each workshop and conference mentioned above. Published contributions come from many sources, conference and workshop proceedings, journal articles, special issues of *IEEE Multimedia, Cutter IT Journal, IEEE Software* and *IEEE Internet Computing,* and the edited book on Web Engineering. From these discussions, It is fair to say that the importance of and need for Web Engineering is now reasonably established, through a consensus among experts on the major differences in the characteristics of Web applications and conventional software[13, 14, 15, 16]. As the authors note, these differences do not arise simply due to the fact that many, early Web developers came from non-software engineering background but because of the new types of (Web) applications. They have all commented on the similarities in application development problems when software engineering was first proposed and the present time in relation to Web development.

Table 1: Major Differences between Web Applications and Conventional Software

| 1. compressed development schedules | 2. constant evolution with shortened revision cycles |
|---|---|
| 3. "content is king", i.e. it is integrated inextricably with procedural processing | 4. insufficient requirement specifications |
| 5. small teams working to very short schedules | 6. emerging technologies/methodologies |
| 7. lack of accepted testing processes | 8. user satisfaction and the threat from one's competition |
| 9. minimal management support | 10. criticality of performance |
| 11. evolving standards to which Web applications should or must comply, depending on the specific circumstances (for example accessibility standards for government sites or IEEE or W3C standards for technological reasons). | 12. understanding of additional disciplines required for Web applications, such as hypertext, graphic design, information architecture |
| 13. security considerations | 14. legal, social and ethical issues |
| 15. variety of backgrounds of developers | 16. Rapidly evolving implementation environment, encompassing various hardware platforms |

Table 1 summarises the experts' findings with a few additional, distinctive characteristics. It is worth noting that this enumeration is based on the experiences of Web developers that the experts had consulted. Once the differences were identified, the question was raised as to whether current software engineering practices could address them successfully. The consensus was that software engineering was needed but was not enough by itself. For more detailed analysis, consult the references cited.

Two points are worth elaborating. They both arise from the *raison d'etre* of the Web, viz. communicating information on a global scale. Table 1 includes them. The first is the nature of information ("content is king") and its effect on the development of a Web application. The second one is the nature of end-users ("user satisfaction and threat of competition") of Web applications.

With regard to information, information systems until now have dealt with largely transactional data in predominantly numerical form, with a bit of textual information, which can be more easily normalised, structured, sorted and searched. Web-based information systems contain text and multimedia, which are difficult to structure, cannot be normalised and are very hard to sort and search. Furthermore, they mix document-orientation with database access through the hypertext metaphor. As content, they are at this time 'integrated inextricably with procedural processing' (part of 'content is king', above). Furthermore, they raise questions of information ownership, and are mired in matters of legal, ethical, social and legal issues[13]. Software developers did not deal with these issues in the past. Web developers must take them into account in creating Web applications. The implication is that if proper policies and procedures are not created, the work of Web developers may not achieve what the client wants[10].

Regarding the nature of the end-users, Web applications may address users anywhere in the world. Unlike the systems in use until now, the Web-based ones are not always confined to specific user groups within an organisation. If Web applications were limited to intranets, the difficulties in understanding the users would be minimised. If they go beyond intranets, however, strategies and policies must be developed to better understand the potential, unknown, and perhaps unknowable, users to establish the quality parameters of the applications in order to deliver quality systems, test sites and applications and maintain security.

## 2.3 Characteristics and Complexity of Web Applications

Web applications vary widely: from small-scale, short-lived services to large-scale enterprise applications distributed across the Internet and corporate intranets. Over the years, Web applications have evolved and become more complex – they range from simple, read-only applications, to full-fledged information systems. This complexity may be in terms of performance (number of hits per second), for example the Slashdot site[17], the Olympics sites receiving hundreds of thousands of hits per minute[18], or in terms of dynamic nature of information, the use of multimedia or in other ways. They may provide vast, dynamic information in multiple media formats (graphics, images and video) or may be relatively simple. Nevertheless, they all demand balance between information content, aesthetics and performance. Table 2, below, brings out the characteristics of early, simple Web-based systems and current, advanced Web-based systems[19].

## 2.4 Multidisciplinary Nature of Web Development

Web applications handle information in its myriad forms (text, graphics, video, audio). Information sciences, multimedia, hypermedia and graphic design deal with structuring, processing, storing and presenting this information. Human-computer Interaction (HCI) and requirements engineering are essential to understand users and their requirements. Network management, general computing and simulation and modelling are required to deliver the information and desired functionality with an acceptable performance level. Software engineering, including new development methodologies, is essential for project and process management.  Since information is very often published for worldwide access, publishing paradigm, and legal, social and ethical issues have to be taken on board. Consequently, good Web development must utilise relevant parts of all these disciplines and not be dominated by narrow viewpoints. Web Engineering is a response in recognition of this multidisciplinary nature of Web applications.  Interestingly, the ACM Computing Curricula 2001 formulates its first principle with a similar statement by stating that "Computing ... extends well beyond the boundaries of computer science"[20].  However, their recommendations cover the entire computing area whereas Web Engineering concentrates on Web development.

| **Table 2: Characteristics of Simple and Advanced Web Applications** | |
|---|---|
| **Simple Web-based systems** | **Advanced Web-based systems** |
| ▪ Primarily textual information in non-core applications<br>▪ Information content fairly static<br>▪ Simple navigation<br>▪ Infrequent access or limited usefulness<br>▪ Limited interactivity and functionality<br>▪ Stand alone systems<br>▪ High performance not a major requirement<br>▪ Developed by a single individual or by a very small team<br>▪ Security requirements minimal (because of mainly one-way flow of information)<br>▪ Easy to create<br>▪ Feedback from users either unnecessary or not sought<br>▪ Web site mainly as an 'identity' for the current clientele, and not as a medium for communication | ▪ Dynamic Web pages because information changes with time and users' needs<br>▪ Large volume of information<br>▪ Difficult to navigate and find information<br>▪ Integrated with database and other planning, scheduling and tracking systems<br>▪ Deployed in mission-critical applications<br>▪ Prepared for seamless evolution<br>▪ High performance and continuous availability is a necessity<br>▪ May require a larger development team with expertise in diverse areas<br>▪ Calls for risk or security assessment and management<br>▪ Needs configuration control and management<br>▪ Necessitates project plan and management<br>▪ Requires a sound development process and methodology<br>▪ User satisfaction vital<br>▪ Web site/application as the main communication medium between the organization and users |

## 2.5  Summary

Web applications are multidisciplinary.  They are built in a constantly changing environment  where requirements are unstable and the development teams typically small.  The user community is wider than before and competition may be spread across the world. Quality Web applications need to be usable, functional, reliable, maintainable, scalable and secure.  These demands on Web applications are radically different from those made on conventional applications.  There is thus a strong need for Web Engineering.

## 3    Evolution and Taxonomy of Web Applications

Web development within an organisation depends upon several factors.  The motivation depends upon the initial purpose of using the Web (Web 'presence' or becoming a Web-based organisation), the customers' expectations and the competitive environment ('keeping up with Joneses')[8].  The drive to systematise development is subject to overall perception of the Web, as depicted in figure 1, and conscious policy decisions within the organisation. For example, a low level perception of the Web is likely to lead to ad hoc, sporadic efforts[21, 22].

As a starting point in understanding the problem domains that the Web currently can address, Table 3 presents a taxonomy of Web applications updated after Ginige and Murugesan[19].  The order of these categories roughly illustrates the evolution of Web applications.  Organisations that started their Web development early may also have followed a similar order in the past.  Although, it is possible to start Web development with applications in any category, this table has been useful to explain to organisations with modest presence on the Web how they might improve or benefit from incremental exposure, thus keeping the risks to the minimum[12].

| *Table 3:  Categories of Web Applications* | |
|---|---|
| ***Category*** | *Examples* |
| ▪ Informational | Online newspapers, product catalogues, newsletters, service manuals, classifieds, e-books |
| ▪ Interactive<br>  ❑ User-provided information<br>  ❑ Customized access | Registration forms, customized information presentation, games |
| ▪ Transaction | E-shopping, ordering goods and services, banking |
| ▪ Workflow | Planning and scheduling systems, inventory management, status monitoring |
| ▪ Collaborative work environments | Distributed authoring systems, collaborative design tools |
| ▪ Online communities, marketplaces | Chat groups, recommender systems, marketplaces, auctions |
| ▪ Web Portals | Electronic shopping malls, intermediaries |
| ▪ Web Services | Enterprise applications, information and business intermediaries |

The take-up of Web technologies and applications within an organisation will not necessarily follow the way the Web has evolved.  Specifically, the World Wide Web was created to solve a specific problem of disseminating information. However, it opened up a novel way of communication and the developers stretched the technologies to make the applications interactive, forcing, in turn, further, rapid innovations in technologies.  It also spawned a new client-server architecture that has become the environment of choice for many applications, both in the Internet and in intranets. The result of this leap-frogging is a wide variety of Web applications, technologies, tools, techniques and methods. The Web is now used to deal with problems in many domains, traditional as well as completely new. An organisation may, therefore, start its own Web development anywhere in the spectrum outlined in Table 3.  The need for Web Engineering has been argued above but how successful it is in delivering Web applications to a satisfactory level will be contingent upon matching the problem domains properly to solution methods and the relevant mix of technologies.  In that sense, Web Engineering is essentially about problem solving.

Web's legacy as an information medium rather than an application medium leads some people to regard Web development primarily as an authoring and publishing problem, giving rise to statements such as "Web development is an art" or it is only "media manipulation and presentation."  The first category of Web applications, 'informational' may seem to fall in this domain, although even with them it will be erroneous to underestimate the total effort and the need to systematise the development work. In any case, this taxonomy should clarify to both organizations and developers that there is a far greater range of Web applications and help them to map out a strategy for Web development within an organisation[20].

As mentioned above, one of the tasks in Web Engineering is to match appropriate methods, technologies (and tools and techniques) to each of these domains.  This is a non-trivial exercise and will be dealt with in a future paper.

## 4    Practice and Research Issues in Developing, Testing and Maintaining Web Applications (Web Engineering)

Section 2 clarified the need for Web Engineering on the basis of collective experience of Web developers, the changing characteristics of Web applications and their multidisciplinary nature. Section 3 traced the evolution and taxonomy of Web applications to establish the domains of problems for Web Engineering.  This section discusses 'what' Web Engineering has to do and recommendations from researchers on 'how' it should be done.

While there are many differences between Web development and Software development, as discussed in section 2, there are also similarities between them.  These include:
- need for methodologies,
- requirements elicitation,
- programming,
- testing, and
- maintenance of those parts that deal with programming and functionalities.

Web Engineering has much to learn from software engineering in these areas but, in the light of the differences enumerated before, software engineering methods may have to be modified or new methods devised.

However, there is one major difference that Web developers/engineers have to bear in mind, as the discussion below clarifies. Web development, and in particular, Web site creation and maintenance, are not merely technical activities. Software development is generally regarded as the province of computing professionals. Web development affects the entire organization, including its interfaces with the world, and has to accommodate non-developers, especially management, when designing or recommending architecture and policies. This is particularly true of content management. Sub-section 4.4 elaborates on this aspect.

### 4.1 Methodologies

A recent survey on Web-based project development by the Cutter Consortium[23], highlights serious problems plaguing large Web-based projects:
- Delivered systems did not meet business needs 84% of the time.
- Schedule delays plagued the projects 79% of the time.
- Projects exceeded the budget 63% of the time
- Delivered system didn't have required functionality 53% of the time.
- Deliverables were of poor quality 52 % of time.

Another survey[24] explicitly reported on the usage of multimedia and Web development techniques and methodologies to suggest that: a) no uniform approach existed, and b) developers need new techniques to do their job.

A caveat is in order. Given the nature of Statistics, the surveys may be statistically valid only for the population they were based on. This is not to question the findings here and a statistical critique of the surveys is beyond the scope of this paper.

At the same time, it is only fair to acknowledge that anecdotal evidence, gathered through personal experience and informal discussions through all the fora cited at the beginning of this paper do point to a wider horizon where the results of both the surveys are borne out. There is also support for this conclusion in the form of absence of a large collection of successful case studies where a systematic approach to Web development was followed and replicated.

Web Engineering has to and aims to improve on this. To this end, several methodologies have been proposed and the experience of their use reported as case studies. For example, see Schwab[25,26] for OOHDM, Ceri et al on WebML[27], Lowe and Henderson-Sellers[28] on OPEN Space Framework, Goeschka and Schranz[29] on their object-oriented engineering framework, Kirda et al[30] on their adaptation of RMM[31] and Conallen[32] on extending the UML. The description and critique of these and other methodologies are beyond the scope of this paper. Section 2 mentioned small teams working to very short schedules as characteristics of Web applications. This is inevitably going to lead to evolutionary approaches to developing such applications. Agile methods and eXtreme programming address similar problems. .

The attraction of the Web and easy availability of development tools have also led to a mushrooming of sites and applications created by end-users, rather than by professionals. The growth in end-user computing logically leads to concerns about the quality and reliability of such applications. Interestingly, but not surprisingly, the Web-based developments have aroused similar concerns from practitioners in other disciplines, who see the Web site and application developers disregarding their traditions and proven techniques (for example, for the concerns of graphic designers, see [33, 34, 35]).

The methodological papers cited before report success of a planned, systematic approach in specific case studies but their repeatability is not fully established nor are their approaches compared with other methods. It is therefore difficult to estimate if the reported successes should be attributed to the generic nature of the proposed methods or to the expertise of the authors who have found methods that work for them.

Engineering is about systematic, disciplined and quantifiable approaches to create usable systems. Among the hallmarks of these approaches are measurability and repeatability of work. Software engineers and other IT professionals lament the fact that software industry is not strong on either. Measurements are scarce and repeatability is exercised more by experience and intuition. There is tremendous opportunity for Web Engineering community to get things right in this arena.

Apart from devising methodologies, a legitimate field of research, Web developers and researchers need to report empirical results of what worked under which circumstances and if it did not, why.

*4.2 Requirements Elicitation*

Insufficient requirements specifications and constant evolution were cited in section 2 as two major differences between Web applications and other software. User-centric approaches and methods to build applications have an unrealised potential in arriving at better specifications. The openness of the Web makes it feasible to get user feedback (and requirements) on-line as opposed to more laborious and expensive traditional methods, such as meetings, interviews, paper-based surveys and focus groups. The on-line methods have not been tried out yet in any great measure and could prove to be very interesting. It is also likely that users now will have a greater say in application development.

Again, Agile methods and eXtreme programming may offer compatible solutions.

*4.3 Testing, Metrics and Quality*

Web testing has many dimensions[11, 36] in addition to conventional software testing. Each unit of a Web application such as page, code, site, navigation, standards, legal requirements must be tested. Usability testing has also become a big and somewhat controversial issue[37]. Services like W3C's HTML, CSS and XHTML certification, and Bobby for accessibility are freely available to Web developers. Consultants in Web site auditing also provide a testing service. However, Web engineers need to create explicit testing strategies that include the relevant tests.

Web metrics and quality are interlinked[38, 39], and like software metrics, under-utilised. However, more tools are becoming available and Web engineers need to evolve conscious policies to test their sites and applications.

*4.4 Maintenance*

Web maintenance, even more than software maintenance, is a continuous activity. Depending on the nature of the application (see section 3, above), the maintenance can become quite complex and does not solely reside in the technical domain. Apart from the evolutionary aspect of Web applications, dealt with in the previous sub-sections, the major differences between Web application and software maintenance arise in relation to content management and site navigation. Content generation, and hence its update and maintenance, will necessarily vary across organisations and applications. The allocation of responsibilities for content may be carried out by human resources (as job descriptions) or other, general management units[40]. Nevertheless, it is imperative that Web developers detail how content maintenance should be carried out and, more importantly, influence the relevant policies.

## 5   Education and Training

The breakneck speed of technological breakthroughs and new types of applications plus the market volatility mean that education and training are now life-long issues for everyone. Web Engineering education and training programmes must therefore not only deal with the current Web technologies but also foster the pro-active approach and a spirit of experimentation and innovation.

Universities organise undergraduate and graduate courses in computing and Information Technology (IT) focussed on traditional disciplines such as Computer Science (CS), Software Engineering (SE), Information Systems (IS) and Computer Engineering and Networking. Web Engineering does not fit well within these boundaries and in fact is not on their radar yet

Professional organisations such as the Association for Computing Machinery (ACM), the Institute for Electrical and Electronics Engineers Computer Society (IEEE-CS), the Association for Information Systems (AIS) and others reinforce the distinctions among these disciplines even as they have started to emphasise commonalities between them. The recent curricula recommendations include net-centric computing subjects at the undergraduate level[41]. However, their thrust is more in terms of technical computing.

On the other hand, there are hundreds of commercial institutions that deliver courses in Web technologies, some of them calling their courses as Web Engineering. They focus mostly on imparting specific skills required for the current technologies and commercial packages. Only infrequently do they prepare students to face a whole gamut of issues, life-long learning or to discharge their social, legal, ethical and professional responsibilities. These areas require a level of maturity, ability and willingness to think through and beyond the boundaries of traditional disciplines.

The School of Computing and IT in the University of Western Sydney introduced two subjects in Web development at undergraduate level in 1997 and a full one-year specialisation at graduate level in 1999. The undergraduate subjects have been very popular, reflecting the popularity of the Web itself. The graduate course currently has about 50 students but there is always a question from practically everyone as to what this Web Engineering means. For details of the curriculum and the principal author's experience in running the courses, see Deshpande et al[42].

Web Engineering is also taught at University of Karlsruhe[43] and has been proposed at University of California, Santa Cruz. See Whitehead[44] for details of the latter course. At the Catholic University in Rio de Janeiro, Web Engineering has been taught, both at the undergratuate and graduate levels, since 1996, with emphasis on design methods. A future paper will carry out a comparative analysis and critique of these courses and other proposals.

We have found that the undergraduate students thought of Web development mainly in technical, computing or programming terms. For them, the new technology itself was fascinating and the rest either irrelevant or a waste of time. Questions of copyright, privacy laws, accessibility, document management, maintenance or how to deal with information explosion were not seen as part of the overall Web development nor were the students mentally quite ready to deal with or devise strategies for future developments. In pedagogical terms, the students are still learning the basics and are not in a position to synthesise that learning, let alone evaluate how they might utilise it.

At the master's level that students start to ask, and can be challenged on how to deal with, the difficult questions about what works in specific situations, what does not, why and what can be done about it. It is at that level, with a certain degree of maturity that they start to devise and systematically experiment with their own solutions.

The tentative conclusion is that Web Engineering is best taught in all its complexity at the graduate level.

## 6   Conclusions

Web Engineering deals with the process of developing, deploying and maintaining Web applications. The main themes of Web Engineering encompass how to successfully manage the diversity and complexity of Web applications development, and, hence, to avoid potential failures that may have serious implications. It is a pro-active approach and at this stage a collection of a body of work. The need for Web Engineering is strong. The task before the Web developers and researchers is to create a robust and tested body of work that can be recommended to suit the specifics of Web applications and environments.